\documentclass[prd,twocolumn,preprintnumbers,nofootinbib]{revtex4}
\usepackage{graphicx}
\usepackage{color}
\usepackage{float}
\newcommand{\rthis}[1]{\textcolor{black}{#1}}
\usepackage{amsfonts,amsmath,amssymb}
\usepackage[plainpages=false, colorlinks=true, anchorcolor=blue, linkcolor=blue, citecolor=blue, bookmarks=false]{hyperref}
\usepackage{natbib}
\usepackage{enumitem}
\usepackage{array}

\pdfoutput=1
\begin{document}
\newcommand{\bthis}[1]{\textcolor{black}{#1}}
\newcommand{\apjl}{Astrophys. J. Lett.}
\newcommand{\apjs}{Astrophys. J. Suppl. Ser.}
\newcommand{\aap}{Astron. \& Astrophys.}
\newcommand{\aj}{Astron. J.}
\newcommand{\araa}{Ann. Rev. Astron. Astrophys. } 
\newcommand{\mnras}{Mon. Not. R. Astron. Soc.}
\newcommand{\ssr}{Space Science Revs.}
\newcommand{\apss}{Astrophysics \& Space Sciences}
\newcommand{\jcap}{JCAP}
\newcommand{\pasj}{PASJ}
\newcommand{\pasp}{PASP}
\newcommand{\pasa}{Pub. Astro. Soc. Aust.}
\newcommand{\physrep}{Phys. Rep.}
\title{A test of Radial Acceleration Relation for the {\it Giles et al Chandra}  cluster sample}
\author{S. \surname{Pradyumna}}\altaffiliation{E-mail:ep18btech11015@iith.ac.in}

\author{Shantanu  \surname{Desai}}  
\altaffiliation{E-mail: shntn05@gmail.com}

\begin{abstract}
We carry out a test of the  radial acceleration relation (RAR) for a sample of 10 dynamically relaxed and cool-core galaxy clusters imaged by the Chandra X-ray telescope, which was studied in ~\cite{Giles_2016}. For this sample, we observe that the best-fit RAR shows a very tight residual scatter  equal to   0.09 dex.  We obtain an acceleration scale of $1.59  \times 10^{-9} m/s^2$, which is about an order of magnitude higher than that obtained for galaxies. \rthis{Furthermore, the best-fit RAR parameters differ from   those estimated from  some of the previously analyzed cluster samples, which indicates that the acceleration scale found from the RAR could  be of an emergent nature, instead of a fundamental universal scale.} 
\end{abstract}
\affiliation{Department of Physics, Indian Institute of Technology, Hyderabad, Telangana-502285, India}
\maketitle

\section{Introduction}
Three recent works~\cite{Tian,Chan20,our_rar_2020} have studied disparate samples of galaxy clusters, in order to ascertain if they  obey the Radial Acceleration relation (RAR). The RAR is a \rthis{tight correlation} between the baryonic ($a_{bar}$) and the total dynamical acceleration ($a_{tot}$). A first definitive case for the RAR was asserted  using the SPARC sample consisting of spiral galaxies~\cite{McGaugh16}, who showed the RAR is obeyed with a scatter of 0.13 dex, with most of this scatter been attributed to observational uncertainties~\cite{Li18}.  The RAR, which was deduced from the SPARC sample can be written as follows:
\begin{equation}
a_{tot} = \frac{a_{bar}}{1-e^{-\sqrt{a_{bar}/a_0}}},
\label{eq:RAR}
\end{equation}
where $a_0 \sim 1.2 \times 10^{-10} m/s^2$~\cite{McGaugh16}.  \rthis{However, some objections to $a_0$ been a  fundamental universal  constant (valid for all galaxies) have been raised~\cite{Rodrigues,delpopolo18,Rodrigues20,Marra}. Ref.~\cite{Rodrigues} showed that the presence of a fundamental acceleration scale  is ruled out at more than 10$\sigma$, and there 
is no  value of $a_0$ which is compatible within 5$\sigma$ for all galaxies. This work concluded that the RAR   is an emergent one, which arises  after stacking the  data from a large number of  galaxies. There is an ongoing debate on these issues~\cite{McGaughNature,Kroupa,delpopolo18,Rodrigues20,Marra}. Most recently, the results in ~\cite{Rodrigues} were also confirmed in ~\cite{Chang,Marra}. (See also Ref.~\cite{Zhu20}).}

More generally,  the RAR  can be re-written  as  a linear regression relation between the logarithms of $a_{bar}$ and $a_{tot}$, $\ln (a_{tot}) = m \ln (a_{bar})+ b$.  All RAR studies for clusters have been obtained by carrying out this linear regression between these two accelerations in logarithmic space~\cite{Tian}.
This relation can be trivially deduced from Milgrom's  MOND paradigm~\cite{LRR}. However, the observed tight scatter cannot be trivially predicted by the standard $\Lambda$CDM model~\cite{Desmond,Navarro,Ludlow}
(see also the discussion in ~\cite{Mcgaugh20}). 
\rthis{However, if $a_0$ is  of emergent nature, as argued in ~\cite{Rodrigues,Marra}, then the RAR could be reproduced  from the complex  interplay between dark matter and baryons~\cite{Ludlow,Navarro,Dutton19}, including the effects due to the  adiabatic contraction of dark in the presence of baryons~\cite{Aseem}.}

We \rthis{have been knowing} for more than three decades that MOND does not work for relaxed galaxy clusters~\cite{White88,Sowmya} (and references therein). Furthermore, relativistic MOND theories cannot explain the coincident GW-EM signal~\cite{Woodard} as well as large-scale structure probes~\cite{Spergel20}. However, a detailed characterization of the RAR in galaxy clusters \rthis{could be an invaluable probe, that could be  used to adjudicate between}  various alternatives to the standard $\Lambda$CDM model, which can reproduce the success of Milgrom's MOND's paradigm, but are consistent with $\Lambda$CDM predictions at larger scales from CMB~\cite{Planck}. 

Motivated by these considerations, three different groups recently carried out a test of the RAR for multiple galaxy cluster samples~\cite{Tian,Chan20,our_rar_2020}.
They have found that the residual scatter between the observed data and best-fit RAR is between 0.11-0.13 dex (see Table I of ~\cite{our_rar_2020} for a summary). However, the aforementioned works have found that the acceleration scale $a_0$ in Eq.~\ref{eq:RAR} is $\mathcal{O} (10^{-9}) m/s^2$, which is about an order of magnitude larger than that obtained for rotationally supported galaxies~\cite{McGaugh16}.
Most recently, this work has also been extended to galaxy groups~\cite{Gopika21}, who find that the scatter is larger, and  $a_0$ falls in-between single galaxies and clusters.

Given the importance of RAR as a probe of $\Lambda$CDM and its alternatives, it is important to cross-check these results with more  cluster samples to gain further insight on how universal this relation is for  clusters, and whether there are any exceptions. In the past three decades a large number of dedicated X-ray based cluster surveys using ROSAT, Chandra, and XMM-Newton (along with other multi-wavelength data) have used the observed cluster samples to address a plethora  of science goals in Cosmology, galaxy evolution, and fundamental Physics ~\cite{Allen,Vikhlininrev}. The same data can also be used for precision tests of RAR.

In this work, we therefore carry out another test of RAR with one such  galaxy cluster sample studied in ~\cite{Giles_2016}, which we refer to as Giles et al cluster sample. This sample has been recently used  to test Verlinde's Emergent gravity paradigm~\cite{Miller20}. The outline of this manuscript is as follows. The basic methodology used for our analysis is described in Sect.~\ref{sec:basic}. The Giles et al cluster sample  along with our data analysis procedure  is outlined in Sect.~\ref{sec:chandra}. Our results are described in Sect.~\ref{sec:results}. We conclude in Sect.~\ref{sec:conclusions}.

\section{Basic Cluster Physics}
\label{sec:basic}
We now provided an abridged description of all the ingredients necessary to carry out a test of RAR using clusters. More details can also be found in our past works~\cite{Gopika,our_rar_2020,Gopika21}.

Galaxy clusters are the largest gravitationally collapsed  objects  in the universe~\cite{Vikhlininrev}. Most of their mass ($\sim 85\%$) comprises dark matter, whereas most of the remaining baryonic mass resides in the form of diffuse hot gas ($\sim 13\%$) in the   Intra-Cluster Medium (ICM). Only a   small fraction of the baryonic mass ($\sim 2\%$)  is  in the form of stars. At X-Ray wavelengths, the hot ICM emits via thermal bremsstrahlung~\cite{Allen}. 
The total mass of a cluster can be estimated from X-Ray measurements assuming that the ICM is in hydrostatic equilibrium within the gravitational potential of the cluster. The total acceleration for test particles in  hydrostatic equilibrium  within a gravitational potential is given by~\cite{Sarazin, Allen}: 
\begin{equation}
a_{tot} (r) = -\frac{k_bT }{r\mu m_p}\left(\frac{d\ln\rho_{gas}}{d\ln r}+ \frac{d\ln T}{d\ln r}\right),
\label{eq:acc}
\end{equation}
Here, $a_{tot}$ is the dynamic acceleration of the cluster as a function of radius from the cluster center ($r$), $k_b$ is the Boltzmann constant, $T$ is the parametrized temperature, $\rho_{gas}$ is the parametrized density of the gas, $\mu m_p$ is the mean molecular weight. We use a $\mu$ value of 0.6 resulting in $\mu m_p  \approx 10^{-27} \,\mathrm{kg}$.
The hydrostatic equilibrium equation is  applicable for  relaxed clusters. 
Once we know the total mass we can estimate the total acceleration $(a_t)$ from Newtonian gravity.

Now, for estimating the acceleration due to baryons, we need to estimate the gas mass ($M_{gas}$) and the stellar mass ($M_{star}$). The gas density ($\rho_{gas}$) profiles are estimated and modelled using the observed projected X-ray  emissivity profile. 
Assuming spherical symmetry, we can  derive the gas mass $M_{gas}$ using:
 \begin{equation}
M_{gas} = \int 4 \pi r^2 \rho_{gas}(r) dr \label{eq:mgas}
\end{equation}

It should be noted that the aforementioned calculations of gas and total mass  assume spherical symmetry. The uncertainty associated with this assumption is about 5\% (See~\cite{Gopika} and references therein).
For stellar mass, we use the  empirical relation obtained in ~\cite{ytlin}, which gives the stellar mass at $R_{500}$, where  $R_{500}$ is the radius at which the overdensity is 500 times the critical density of the universe, i.e. $\rho (R_{500})= 500 \left( \frac{3H^2}{8 \pi G} \right)$. The total star mass $M_{star} (R_{500})$ at $R_{500}$ is given by: 

\begin{equation}
    \frac{M_{star} (R_{500})}{10^{12} M_{\odot}} = 1.8 \left(\frac{M_{500}}{10^{14} M_{\odot}}\right)^{0.71}
\label{eq:mstar}
\end{equation}

Then, to estimate the stellar mass at any  other radius ($r$), we assume that the stellar component  follows an isothermal distribution, and we thereby obtain~\cite{Rahvar}:
\begin{equation}
    M_{star}(r)=\left(\frac{r}{R_{500}}\right)M_{star}(R_{500})
    \label{eq:mstarr}
\end{equation}
The baryonic acceleration ($a_b$) can then be obtained  at any value of the radius $r$ from the sum of  the gas and star mass.

\section{Data and Analysis:}
\label{sec:chandra}
Giles et al~\cite{Giles_2016} (G17, hereafter) provide the temperature and  gas density profile fits for 34  
clusters within the  redshift range $0.15\le z \le 0.3$, which were imaged with the  Chandra X-ray telescope. The main goal of this work was to characterize the scatter in the Luminosity-Mass relation for this sample.
This cluster sample was selected from the ROSAT based Brightest cluster sample survey~\cite{Ebeling98,Ebeling00} with  a lower luminosity cutoff of $L_X$ (0.1-2.4 keV) $\geq 6 \times 10^{44}$ erg/sec~\cite{Dahle}. This cut provided a sample of 36 clusters. From these, A689 was culled because of data quality issues and Zw5768  did not pass the luminosity cut using  a  revised redshift estimate. The exposure times for these clusters are in the range of 7.3-450 ks (cf. Table 1 of G17).
The luminosity of these clusters in X-Ray band of 0.1-2.4 keV falls  between $\sim (6.3-24.5)\times 10^{44} $ erg/s. More details of the observations and   data  reduction can be found in G17. 

Since  hydrostatic equilibrium can  be robustly  used to obtain the mass only if the cluster is dynamically relaxed and has a cool core, we select a sub-sample of relaxed as well as cool core clusters for our analysis. This was done in two steps: (1) the clusters were determined to be in a relaxed or non-relaxed state, on the basis of  the centroid shift ($\langle w \rangle$)~\cite{Poole}, the description of which can be found G17; (2) the cluster sample was further bifurcated into  cool core and non-cool core sample. A cut of $\langle w \rangle < 0.009 r_{500}$ was made to select dynamically relaxed clusters.  The presence of a cool core  was based on the calculating  the central cooling time given by: $$t_{cool} (yr) =8.5\times 10^{10} (\frac{n_p}{10^{-3}})^{-1} (0.079kT_{CCT})^{1/2}$$
The last parameter used for the selection of cool core sample was  the cuspiness, which is  defined as the logarithmic slope of gas density at $0.04r_{500}$~\cite{Hudson}. In summary, cuts were applied on $\langle w\rangle<0.009 r_{500}$, $t_{cool}< 7.7 Gyr$, cuspiness $>0.7$, to select the dynamically relaxed and cool core sample. 
After applying these cuts, we are left with a  sample of  10 clusters (from the original dataset of  34 clusters) for our analysis.
They are: A2204, RXJ1720.1+2638, A1423, Z2089, RXJ2129.6+0005, A1835, MS1455.0+2232, RXJ0439.0+0715, RXJ0437.1+0043, Z3146.

The gas density ($\rho_{gas}$) model used  is equal to  $1.624 m_p\sqrt{n_p(r)n_e(r)}$, where $m_p$ is the proton mass; $n_p$ and $n_e$ are the proton and electron number densities, respectively. Their product ($n_e n_p$) is parametrized by a modified single-$\beta$ profile:
\begin{equation}
n_e (r)n_p(r) = \frac{(r/r_c)^{-\alpha}}{(1+r^2/r_c^2)^{3\beta -\alpha /2}}\frac{n_0^2}{(1+r^\gamma/r_s^\gamma)^{\epsilon/\gamma}}
\label{eq:gasdensity}
\end{equation}
This is a variant  of the double-$\beta$ profile used by Vikhlinin et al~\cite{Vikhlinin06}, which was also used in our previous works~\cite{Gupta1,Gupta2,Gopika,our_rar_2020} The second term in the  double-$\beta$ profile was dropped so as to fit both the low and high quality data in the sample.

The temperature profile used for our current analysis is:
\begin{equation}
T(r) = T_0 \frac{(x+T_{min}/T_0)}{x+1}\frac{(r/r_t)^{-a}}{\left[1+(r/r_t)^b\right]^{c/b}},
\label{eq:temp}
\end{equation}
where $x=\left(\frac{r}{r_{cool}}\right)^{a_{cool}}$. The   values of the free parameters in Eq.~\ref{eq:temp} for these clusters can be found in    Table A2 of G17.
The error estimates  for the temperature and density have been obtained in G17  using Monte-Carlo simulations.
For this work, the temperature error bars used in our analysis have been obtained by digitization of the plots in G17, which in turn were used to propagate the errors in $M_{tot}$. 
For the errors on the gas density, we again digitized the emission measure plots given in G17. We calculated the fractional errors in the emission measure data at the radii of interest by using \textit{spline} interpolation. Then using the fact that the gas density is proportional to the square root of the emission measure, the fractional error in the density is given as: $\frac{\Delta  \rho_{gas}}{\rho_{gas}}=\frac{1}{2} \frac{\Delta EM}{EM} $ where EM stands for the emission measure value.  
We now plug in the values for $\rho(r)$ and $T(r)$ from Eqns.~\ref{eq:gasdensity} and \ref{eq:temp} respectively in Eq.~\ref{eq:acc} to calculate the total dynamical  acceleration $a_{tot}(r)$. 

To estimate baryonic mass, we need to determine the  gas and star mass. The gas  mass can be  estimated from the gas density profile  in  Eq.~\ref{eq:gasdensity}. 
To estimate the  stellar mass,  we need $R_{500}$ and $M_{500}$ for these clusters. G17 provides these $R_{500}$  estimated using the hydrostatic mass estimates. We plugged  these values in Eq.~\ref{eq:mstar} to obtain the total stellar mass estimate at $R_{500}$. The stellar mass at all other radii was obtained from Eq.~\ref{eq:mstarr}.
Therefore, the baryonic acceleration can  then estimated from this total baryonic mass assuming Newtonian gravity.

Now, we do a joint fit to the RAR using the calculated data along with associated errors for $a_{tot}$ and $a_{bar}$ at $r$ = 100, 200, 400, and 600 kpc for all the clusters. The only exception is Z2089, for which we used measurements at  100, 200, 400, and 560 kpc.   Similar to previous works~\cite{Tian,our_rar_2020}, we do a fit in log-acceleration space by doing a linear regression to the relation: 
\begin{equation}
y=m x + b
\label{lineareq}
\end{equation}
where $y=\ln (a_{tot})$ and $x=\ln (a_{bar})$. 
We then  maximize a log-likelihood  given by :
\begin{eqnarray}
-2\ln L &=& \large{\sum_i} \ln 2\pi\sigma_i^2 + \large{\sum_i} \frac{[y_i-(mx_i+b)]^2}{\sigma_i^2}
\label{eq:eq13}  \\         
\sigma_i^2 &=& \sigma_{y_i}^2+m^2\sigma_{x_i}^2+\sigma_{int}^2
\label{eq:error}
\end{eqnarray}
Here, $\sigma_{y_i}$ and $\sigma_{x_i}$ denote the errors in $y$  and $x$ respectively; $\sigma_{int}$ is the intrinsic scatter for our linear relation between the logarithms of the accelerations, which is a free parameter. 
We used the {\tt emcee} MCMC sampler~\cite{emcee} to \rthis{sample} the likelihood in  Eq.~\ref{eq:eq13}. The total number of steps and burn-in steps used for the MCMC analysis are 3000 and 500 respectively. The number of walkers used for our MCMC analysis is equal to  40. \rthis{We used uniform non-informative priors on $m$, $b$, and $\ln (\sigma_{int})$ given by: $m \in [-10,10]$ ; $b \in [-100,100]$; and $\ln (\sigma_{int}) \in [-5,-1]$}.

For calculating the residual scatter in dex, we first calculate residuals by subtracting the $\log_{10}(a_{dyn})$ predicted by our best-fit from  $\log_{10}(a_{dyn})$ calculated from the data. This is the same practise used in the literature to characterize the scatter in the RAR~\cite{our_rar_2020}.
We then fit a normal distribution with a mean of zero to these residuals. The standard deviation obtained is used to characterize the  residual scatter in dex.

\section{Results}
\label{sec:results}
\begin{figure}[H]
    \centering
    \includegraphics[width=0.5\textwidth]{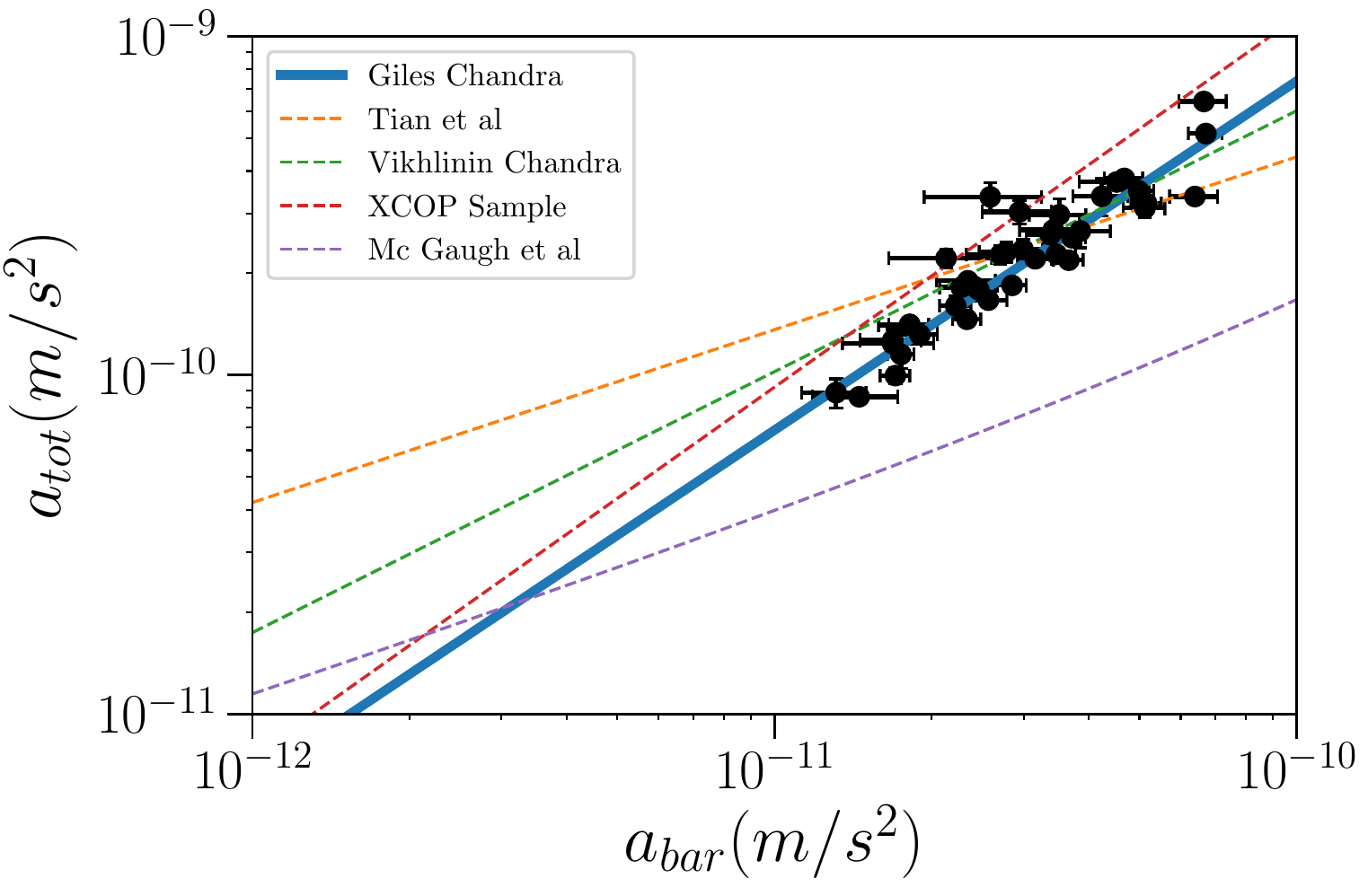}
    \caption{The best-fit RAR(blue solid line) obtained for $a_{tot}$ vs $a_{bar}$ data for a sample of 10 dynamically relaxed and  cool-core galaxy clusters from the G17 catalog. Best fits from previous  works  on RAR for clusters have been overlaid for comparison.}
    \label{fig:giles_rar}
\end{figure}

\begin{figure*}
\centering
    \includegraphics[width=1\textwidth]{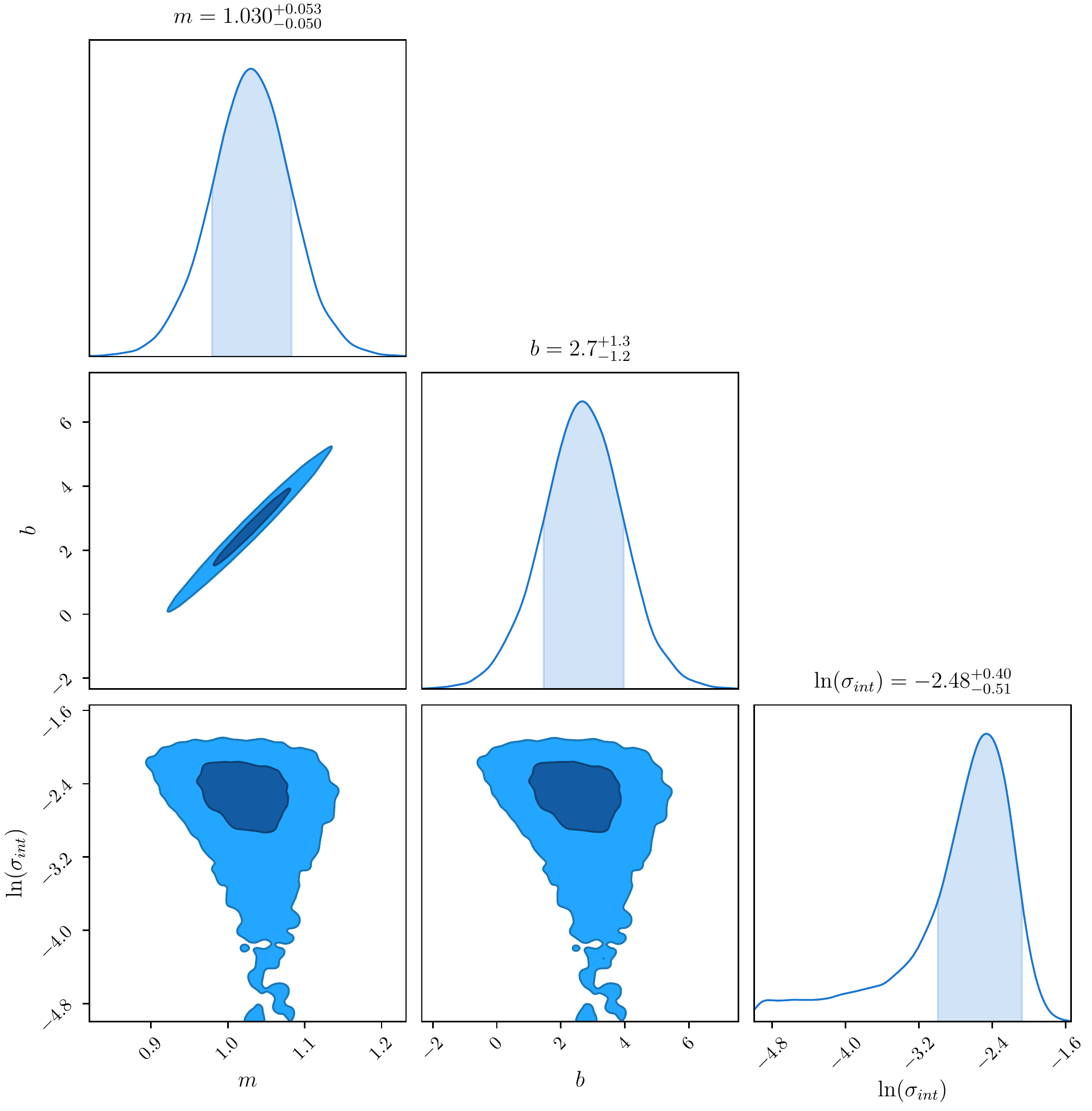}
    \caption{68\% and 95\% marginalized credible intervals obtained for the slope, intercept, and natural log of the  intrinsic scatter on fitting the RAR  to the Giles et al Chandra galaxy cluster sample~\cite{Giles_2016}.}
    \label{fig:giles_corner}
\end{figure*}

We now present the results of our analysis of 10 relaxed cool core clusters from this sample. Fig.~\ref{fig:giles_rar} shows the best-fit line obtained for the data along with the best fits obtained by McGaugh et al~\cite{McGaugh16} (for galaxies), Tian et al~\cite{Tian}, Chan and Del Popolo~\cite{Chan20} and our previous analysis~\cite{our_rar_2020} (on galaxy clusters). 
Fig~\ref{fig:giles_corner} shows the 68\% and 95\% credible intervals for the slope, intercept, and the natural log of intrinsic scatter obtained using the \textit{ChainConsumer} post-processing package~\cite{chainconsumer}, using the MCMC chains from {\tt emcee} as inputs.
The best-fit slope and intercept obtained are $m=1.03 \pm 0.05$ and $b=2.7^{+1.3}_{-1.2}$ respectively. The residual scatter obtained for the best fit is about 0.08 dex. 
The best-fit values and the scatter are shown in Table~\ref{tab:results} alongside the fits obtained in previous works for comparison. 
We note that the best-fit slope and intercept observed for XCOP sample and the current sample agree within $1\sigma$. However, the best-fit parameters of the G17 sample  differ from the other cluster samples, such as the Vikhlinin et al Chandra sample~\cite{Vikhlinin06}, HIFLUGS sample~\cite{Chan20}, and the CLASH sample~\cite{Tian} to between $3-5\sigma$.
The acceleration scale that we obtain for current sample on fixing the slope to $m=0.5$ is  equal to $(1.59 \pm 0.14)   \times 10^{-9} m/s^2$. \rthis{The error in $a_0$ was estimated
using error propagation, based on the $1\sigma$ error in $b$, which was obtained from redoing the RAR fit for a fixed slope of 0.5.} The acceleration scale obtained for this sample of galaxy clusters is therefore, an order of magnitude higher than the $a_0$ obtained from the SPARC sample~\cite{McGaugh16} for galaxies, in agreement with all  previous analyses  carried out  on galaxy clusters.~\cite{Tian, Chan20, our_rar_2020}. 

When we try to apply a cut to look for dynamically relaxed but non-cool clusters, using only the dynamically relaxed criteria mentioned in Sec~\ref{sec:chandra}, we find three additional  clusters for our analysis. 
Then, we tried to do a similar analysis on all the relaxed clusters (both cool core and non-cool core).
The best fit slope and intercept obtained in this case are $m = 0.92 \pm 0.07$ and $b = 0.0 \pm 1.7$, and the scatter increases slightly to 0.095 dex.

Therefore, our results reinforce those in our previous work, and indicate the RAR is obeyed for clusters with a tight intrinsic scatter. However, this relation is not universal, as all  the  cluster samples
do not agree on the RAR best-fit parameters. \rthis{These inconsistent fits   among the different cluster samples, reaffirm the view that $a_0$ could be   of  emergent nature, instead of a fundamental acceleration scale, as previously discussed~\cite{Rodrigues,Marra}.}

\begin{table*}[t]

\begin{tabular}{|l|l|l|l|l|l|}
\hline
Cluster Sample         & Slope                     & Intercept                    & Intrinsic scatter          & $a_0$ $(m/s^2)$                  & Scatter (dex) \\ \hline
Giles Chandra  (this work) & $1.03^{+0.05}_{-0.05}$    & $2.7^{+1.3}_{-1.2}$       & $0.07_{ - 0.05}^{ + 0.03}$ &                $(1.59 \pm 0.14) \times 10^{-9}$                   & 0.08          \\ 
XCOP Sample~\cite{our_rar_2020}           & $1.09^{+0.07}_{-0.07}$    & $4.44^{+1.80}_{-1.85}$       & $0.21_{ - 0.02}^{ + 0.03}$ & $(1.12 \pm 0.11) \times 10^{-9}$ & 0.11          \\ 

Vikhlinin Chandra~\cite{our_rar_2020} & $0.77\pm 0.1$& $-3.5^{+2.6}_{-2.7}$&($0.0002 \pm 0.018$)\% & $(9.26 \pm 1.66) \times 10^{-10}$ & 0.14\\
Chan et al~\cite{Chan20}             & $0.72_{- 0.06}^{ + 0.06}$ & $-5.49 _{- 1.39}^{ + 1.39 }$ & $0.27_{-0.02}^{+0.02}$     & $9.5 \times 10^{-10}$            & 0.18          \\ 
Tian et al~\cite{Tian}             & $0.51^{+0.04}_{-0.05}$    & $-9.80^{+1.07}_{-1.08}$      & $14.7^{+2.9}_{-2.8}\%$      & $(2.02\pm 0.11) \times 10^{-9}$  & 0.11          \\ \hline
\end{tabular}
\caption{The best-fit values obtained for the Chandra sample of G17 along with those obtained for other cluster samples. We see that the acceleration scale we obtain from this sample is in agreement with results from previous RAR analysis with clusters and about an order of magnitude higher than that obtained for galaxies.}
\label{tab:results}
\end{table*}

\section{Conclusions}
\label{sec:conclusions}
We implement a test of the scatter in the  correlation between the total dynamical and baryonic acceleration in log-log space (dubbed as RAR) using a sample of 10 dynamically relaxed cool core clusters,  obtained from a dataset of  35 ROSAT selected clusters imaged with Chandra, and described in  detail in G17. The aim of this work was to further assess the conclusions of previous studies of RAR using  ancillary cluster samples~\cite{Tian,our_rar_2020,Chan20}. These previous analyses showed that  the intrinsic scatter  of the RAR is the same as that seen for spiral galaxies~\cite{McGaugh16}, but the observed acceleration scale is elevated by an order of magnitude, and the best-fit values are inconsistent among the cluster samples~\cite{our_rar_2020}.

The best-fit values for the parameters of the RAR using the G17 cluster sample can be found in Fig.~\ref{fig:giles_corner}. The best-fit RAR along with the data is shown in Fig.~\ref{fig:giles_rar}. A comparison of our best-fit results along with those from the previous works is summarized in Table~\ref{tab:results}. We find that the scatter for this sample is equal to 0.08 dex and is slightly increased to 0.094 dex, once we include the relaxed non-cool core clusters. Our estimate for the acceleration scale  we obtained by positing a slope of 0.5 in the RAR regression relation is about  about an order of magnitude larger than that deduced using the SPARC sample.
The best-fit value for the slope and intercept is in agreement with the  same for the XCOP sample, but differ from other cluster samples previously analyzed by upto $5\sigma$. This shows that the RAR is not universal for clusters. \rthis{Therefore, this   reinforces the fact alluded to in ~\cite{Rodrigues,Marra}, that the acceleration scale deduced from RAR  is not a fundamental universal constant, but instead  could be an emergent effect caused by stacking data from different clusters.}

\begin{acknowledgements}
We would like to thank Paul Giles for useful correspondence about G17. \rthis{We are also grateful to the anonymous referee for useful  feedback on the manuscript.}
\end{acknowledgements}
\bibliography{references}
\end{document}